\documentclass{article}
\usepackage{spconf,amsmath,graphicx,amssymb,subfig}
\usepackage{xcolor}
\usepackage{soul}
\usepackage{hyperref}
\usepackage{geometry}
\hypersetup{
    colorlinks=true,
    linkcolor=blue,
    filecolor=magenta,      
    urlcolor=cyan,
    pdftitle={Overleaf Example},
    pdfpagemode=FullScreen,
    }

\title{LOW-LATENCY SPEECH ENHANCEMENT VIA SPEECH TOKEN GENERATION}

\name{Huaying Xue, Xiulian Peng, Yan Lu}
\address{Microsoft Research Asia, Beijing, China}
\begin{document}
%
\maketitle

\begin{abstract}
Existing deep learning based speech enhancement mainly employ a data-driven approach, which leverage large amounts of data with a variety of noise types to achieve noise removal from noisy signal. However, the high dependence on the data limits its generalization on the unseen complex noises in real-life environment.   
In this paper, we focus on the low-latency scenario and regard speech enhancement as a speech generation problem conditioned on the noisy signal, where we generate clean speech instead of identifying and removing noises. Specifically, we propose a conditional generative framework for speech enhancement, which models clean speech by acoustic codes of a neural speech codec and generates the speech codes conditioned on past noisy frames in an auto-regressive way. Moreover, we propose an explicit-alignment approach to align noisy frames with the generated speech tokens to improve the robustness and scalability to different input lengths. Different from other methods that leverage multiple stages to generate speech codes, we leverage a single-stage speech generation approach based on the TF-Codec neural codec to achieve high speech quality with low latency. Extensive results on both synthetic and real-recorded test set show its superiority over data-driven approaches in terms of noise robustness and temporal speech coherence. 
\end{abstract}
%


\section{Introduction}
\label{sec:intro}

\begin{figure}[tb]
\centering

\subfloat[Mainstream DNN-based framework]{
	\label{subfig:framework_traditional}
	\includegraphics[width=0.5\textwidth]{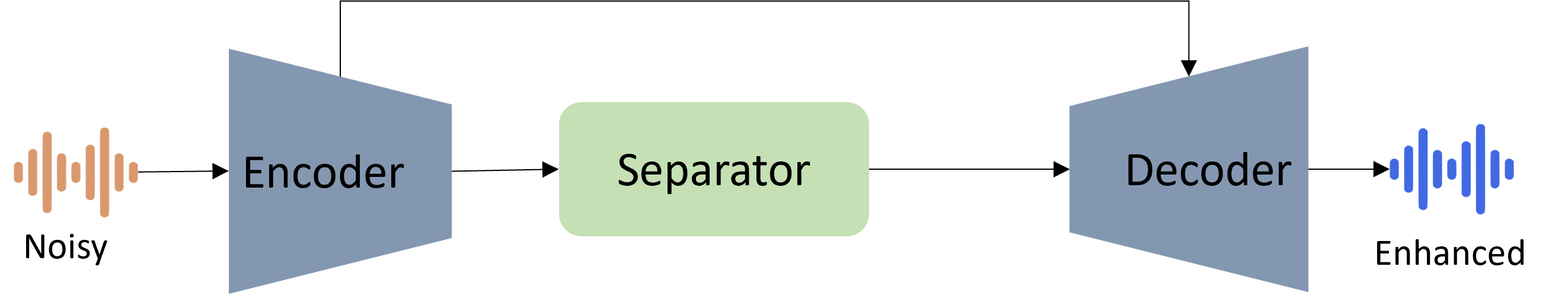} } 

\subfloat[Proposed generative framework]{
	\label{subfig:framework_generative}
	\includegraphics[width=0.5\textwidth]{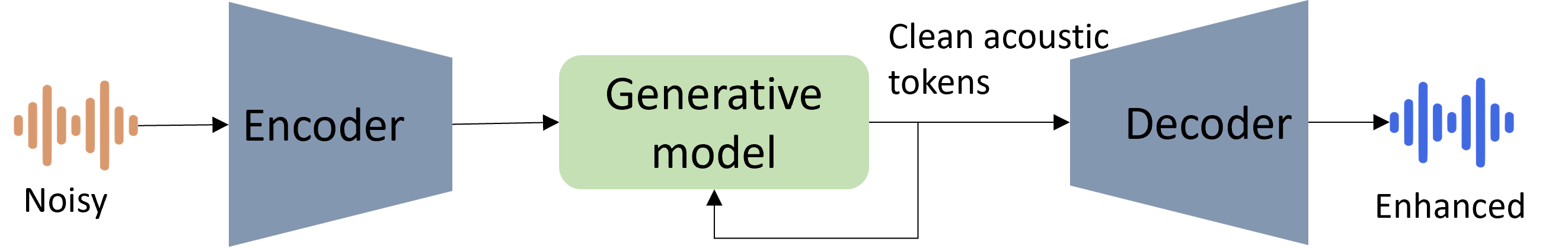} } 
	
\caption{The proposed generative framework for speech enhancement.}
\label{fig:framework_comparison}

\end{figure}

Recently speech enhancement via deep learning methods has achieved great success in real applications, such as real-time speech communications. Existing deep learning based methods mainly leverage large amounts of data with a variety of common noise types to achieve fairly good performance on general cases. They try to separate speech and noise signals from the noisy one with a regression model by an end-to-end supervised learning, as shown in Fig. \ref{subfig:framework_traditional}. 
Prior works \cite{luo2019conv,luo2020dual,zhang2020furcanext} operate in the time domain and predict the target clean speech from the waveform of the noisy mixture directly. Other methods instead operate in the time-frequency domain \cite{zhao2016dnn,yin2020phasen}, using either mapping-based targets by the complex spectral mapping (CSM) \cite{tan2019complex} or masking-based targets, e.g. spectral magnitude mask (SMM) \cite{wang2014training}, phase-sensitive mask \cite{erdogan2015phase}, and complex ratio mask (CRM) \cite{williamson2015complex} that takes phase into consideration for a good reconstruction. A typical U-Net based framework is leveraged to extract and reconstruct features and a long-context temporal modeling module is utilized for separation by a temporal convolutional network (TCN) \cite{luo2019conv} or recurrent networks \cite{weninger2015speech}. Such data-driven approaches still cannot generalize well to unseen and complex noise types in the real-life environment.

 \begin{figure*}[htb] 
\centering 
\includegraphics[width=1.0\textwidth]{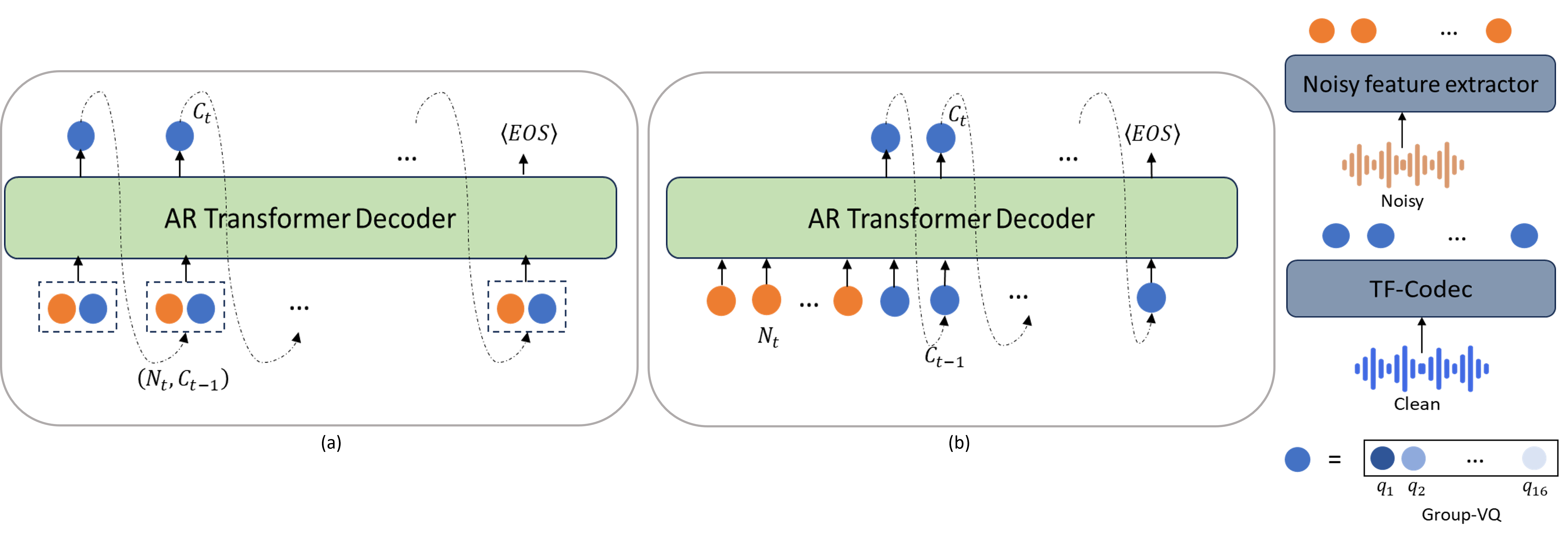} 
\vspace{-0.6cm}
\caption{The proposed framework with the explicit-alignment scheme for generation. The noisy token is extracted by a noisy feature extractor. The clean token is extracted from the pre-trained TF-Codec and managed in a group-VQ manner. (a) The proposed explicit-alignment based generative model. (b) A typical prefix-based conditional generative model.}
\vspace{-0.6cm}
\label{Fig_proposed_framework} 
\end{figure*}
Some recent works tackle the speech enhancement problem with speech synthesis technologies. These methods usually predict the intermediate features of the clean speech and reconstruct the waveform with a vocoder, e.g. HiFi-GAN \cite{su2020hifi}. \cite{polyak2021high} predicts several disentangled speech representations like content features, speaker feature, \(F_0\) and loudness feature, while \cite{su2021hifi} predicts the MFCCs of the clean speech. And \cite{irvin2023self} directly predicts the clean speech through the vocoder from a noisy representation extracted from a large self-supervised learning (SSL) model. These methods model speech with prior knowledge and rely on a strong vocoder to achieve good synthesis quality. However, they mostly work for the offline scenario with long latency.

In this paper, we focus on the low-latency scenario and formulate speech enhancement as a speech generation problem. We propose a conditional generative framework which models clean speech by neural speech codec codes, and predicts the speech code of each frame conditioned on past and current noisy input in an auto-regressive way, as shown in Fig. \ref{subfig:framework_generative}. Particularly, we propose an alignment scheme to transform the generative model into a purely next-token prediction approach without any condition, with which the robustness and scalability to different input lengths are largely boosted. Moreover, unlike other speech generation methods \cite{borsos2023audiolm, wang2023neural} that leverage neural codec with a residual vector quantization (RVQ) to generate speech with multiple prediction stages and non-causal condition, we take the TF-Codec \cite{jiang2022end, jiang2023latent}, a high-quality low-latency speech codec with parallel group quantization, to model speech. In this way, we can generate acoustic token in a single stage with low latency and still achieve high quality. Extensive results on both synthetic and real-recorded test sets show its superiority over data-driven methods in terms of noise robustness and less speech distortion. Ablation studies on variable lengths of audios also demonstrate the effectiveness of our alignment scheme. 

The remainder of this paper is organized as follows. Section~\ref{sec:proposed framework} describes the proposed framework in detail. Section~\ref{sec:experimental results} presents the experimental settings, the comparison with data-driven methods and various ablation studies. Finally, Section~\ref{sec:conclusions} concludes our paper.

\section{The proposed framework}
\label{sec:proposed framework}
\subsection{Overview}
\label{ssec:overview}
We focus on low-latency speech enhancement in this paper. Given time-frequency domain input of the noisy signal \(\mathrm{X} = \{x_{1},x_{2},...,x_{T}\}\), where $x_t,t=1,2,..,T$ is the $t$-th frame and $T$ is the number of frames, and that of the clean target \(\mathrm{Y} = \{y_{1},y_{2},...,y_{T}\}\), we formulate the low-latency speech enhancement task as a conditional generative modeling problem given by
\begin{equation}\label{overview}
    P(\mathrm{Y}|\mathrm{X}) = \prod \limits_{t=1}^Tp(y_t|y_{<t},x_{\leq{t}}).
\end{equation}

To model clean speech \(\mathrm{Y}\), we leverage the clean codes from a pretrained neural codec TF-Codec. With this, each frame of the clean speech will be encoded into discrete acoustic codes, denoted by \(\mathrm{C}^{T\times{K}} = \{C_{1},C_{2},...,C_{T}\}\), where each frame $C_t,t=1,2,..,T$ has \(K\) codes from \(K\) parallel group quantizers of the neural codec. We employ a causal noisy feature extractor to get features \(\mathrm{N}^{{T}\times{D}} = \{N_{1},N_{2},...,N_{T}\}\), where \(D\) is the latent dimension of the noisy feature. Then the generative problem evolves into
\begin{equation}\label{overview}
    P(\mathrm{C}|\mathrm{N}) = \prod \limits_{t=1}^Tp(C_t|C_{<t},N_{\leq{t}}).
\end{equation}
With predicted speech tokens, we get can the speech waveform by decoding with the neural codec.

We leverage an auto-regressive transformer decoder for the speech token generation from noisy features. Unlike typical ways to take noisy features as a condition(Fig.\ref{Fig_proposed_framework}b), we design an explicit-alignment approach(Fig.\ref{Fig_proposed_framework}a) to explicitly align the noisy feature with the clean speech codes to be generated in the temporal dimension by binding each noisy frame with the previous generated token. For causal speech generation, we leverage a single-stage token generation based on the TF-Codec codes, where all $K$ codes of each frame $C_t$ are predicted jointly in a single step. The following sections will describe them in detail.


\subsection{Speech tokenizer with TF-Codec}
\label{sssec:speech tokenizer with TF-Codec}
We use the pretrained causal neural speech codec TF-Codec \cite{jiang2023latent} to extract the acoustic token of each clean frame \(\mathrm{C_t}\). Unlike \cite{jiang2023latent}, we remove the predictive loop and use the non-predictive model at 6 kbps for efficient acoustic modeling with high quality output. Specifically, the TF-Codec takes the magnitude-compressed time-frequency spectrum with a window length of 20 ms and a hop length of 5 ms as input. Then a stack of 2D causal convolutional layers, followed by a temporal convolutional module (TCM) and a gated recurrent unit (GRU) block is used to capture the short-term and long-term temporal dependencies between the input frames in a causal manner. For the quantization, it combines 4 frames together, producing a frame rate of 50 Hz for quantization. Instead of using RVQ, it employs group quantization where the latent embedding is splitted into $K$ groups and each group is quantized by a vector quantizer with a codebook of 1024 codewords. All $K$ acoustic codes are concatenated and decoded to get the reconstructed waveform. Here We take acoustic TF-Codec codes at 6 kbps, where each frame feature is encoded by \(K=16\) group vector quantizers at 50 Hz. 

\subsection{Noisy feature modeling}
\label{sssec:noisy feature modeling}
The noisy feature extractor shares the similar structure as the encoder in TF-Codec but it is trained from scratch together with the generative model. To align with the resolution of acoustic codes in the temporal dimension, we also temporally downsample by \(\times{4}\) by simply concatenating 4 neighboring frames in the channel dimension, producing features \(\mathrm{N} = \{N_{1},N_{2},...,N_{T}\}\). It will be linearly projected to \(D_{token}\) channels as the input of the generative model.

\subsection{Generative model with explicit-alignment scheme}
\label{sssec:explicit alignment}
We leverage auto-regressive transformer decoder as the generative model for speech token generation. As shown in Fig.\ref{Fig_proposed_framework}a, to achieve frame-by-frame causal generation, we explicitly align the noisy condition \(N_t\) with the target clean codes to be predicted \(C_t\) by binding \(N_t\) with previous token \(C_{t-1}\) at each time step \(t\). So for each time step, the input of transformer is \((N_t, \mathrm{C}_{t-1}), t\leq{T}\). Specifically, \(N_t\) and \(\mathrm{C}_{t-1}\) are concatenated in the channel-wise. Furthermore, we use cross-attention to calculate the correlation between \(C_t\) and the pairs \(\{(N_t,C_{t-1}), t\leq{T}\}\) in transformer by extracting query from \(C_t\) and key-value from \(\{(N_t,C_{t-1}), t\leq{T}\}\). We find such usage of noisy condition brings much more robust prediction for each frame and can be generalized to longer sequence.

\subsection{Single-stage causal speech generation}
\label{sssec:causal speech generation}
As the group quantization in TF-Codec encodes each group independently, we leverage a single-stage causal speech generation to generate clean codes of all $K$ quantizers simultaneously for each frame. Specifically, the concatenated embeddings corresponding to all $K$ quantizers of the previous token are taken as the input, combined with the aligned noisy feature frame. For each group embedding, the dimension is \(D_{token}/K\). $K$ classification heads are employed to predict the $K$ acoustic codes for current frame.

\subsection{Training algorithm}
We leverage teach-forcing technique to train this generative model with cross-entropy loss given by
\begin{equation*}
    \begin{split}
     &\max{P(\mathrm{C}|\mathrm{N};\theta_{AR})} =\\
&\max{\prod \limits_{t=1}^{T+1}p(\mathrm{C}_t|(N_1,\mathrm{C}_{0}),(N_2,\mathrm{C}_1),...(N_{t},\mathrm{C}_{t-1});\theta_{AR}}),
    \end{split}
\end{equation*}
where \(\mathrm{C}_{0}\) is the $\langle{start}\rangle$ token, \(N_{T+1}\) is the null token which is combined with \(\mathrm{C}_T\) to predict the $\langle{eos}\rangle$ token. $\theta_{AR}$ denote the generative model parameters. During inference, the actual predicted token is used as previous tokens frame by frame.

\section{EXPERIMENTAL RESULTS}
\label{sec:experimental results}
\subsection{Experiment setup}
\subsubsection{Datasets and settings}
\label{sssec:Datasets and Settings}
For training set, we use 580 hours of 16khz training data in LibriTTS as clean English speech, in which train-other has been enhanced with a pretrained speech enhancement model to remove the reverb. For noise and room impulse responses(RIR), we use noise and RIR dataset from Deep Noise Suppression (DNS) Challenge at ICASSP 2022 \cite{dubey2022icassp}. Then we synthesized 500 hours noisy samples with SNR uniformly distributed from -5 to 20dB and speech level ranging from -35 to -15dB. Each sample has 10 seconds. For testing, we use a synthetic test set with multilingual noisy speech and the blind test set with real-recorded samples from DNS Challenge at ICASSP 2022 \cite{dubey2022icassp}. The blind test set is downsampled from 48kHz to 16kHz to meet our scenario. Each test clip has a 10s duration. To verify the intelligibility, the test-clean dataset from LibriTTS synthesized with noise is used for the word error rate (WER) evaluation. The audio length varies from 1s to 15s. After removing some low-activity and clipping audio files, we finally take 1806 files for this evaluation.

The auto-regressive generative model leverages a transformer decoder structure with 6 layers, 8 attention heads, a embedding dimension \(D_{token}\) of 512, feed-forward layer dimension of 2048 and dropout of 0.1.
In transformer, each quantizer has its own embedding space whose dimension is 32. For noisy feature extractor, the kernel size of the 4 convolution layers is \((2,5)\), with the stride of frequency dimension \((1,4,4,2)\) and channel dimension \((16,32,64,64)\). The kernel size of 4 TCM blocks is 5 with dilation rates \((2^0, 2^1, 2^2, 2^3)\) and middle channel dimension of 512. The noisy feature dimension \(d\) is set to 480. Each codebook of TF-Codec has 1024 codewords with a dimension of 120, resulting in a codebook size of 1.875M parameters. The parameters for the decoder and the generative model are 3.33M and 4.5M, respectively. 

The models are trained using 8 NVIDIA TESLA V100 32GB GPUs with a batch size of 64 for 400 epochs. We optimize the models with the Adam optimizer with a learning rate of \(2\times10^{-4}\). For inference, we use temperature sampling with a temperature of 0.8 and choose the best one from 3 inferences.

\subsubsection{Evaluation metrics}
\label{sssec:Evaluation metrics}
To evaluate the DNS performance, we use DNSMOS P.835\cite{reddy2022dnsmos} model to calculate the speech quality (SIG), background noise quality (BAK) and the overall quality (OVR). 
To evaluate the intelligibility of the generated speech, we use the pretrained Automatic Speech Recognition (ASR) model on LibriSpeech from WeNet toolkit \cite{zhang2022wenet} to calculate WER.

\subsubsection{Baseline}
\label{sssec:baseline}
To compare the speech enhancement performance with data-driven approaches, we implement a low-latency model based on TFNet \cite{jiang2022end}, a UNet-style structure, with causal convolutional encoder and decoder and skip connections between them. Interleaved structure of TCMs and G-GRUs \cite{jiang2022end} are used to extract long-term dependencies of input frames. The encoder and decoder are composed of 6 layers with kernel size \((2,7)\), stride of frequency dimension \((1,2,2,2,2,2)\), channel dimension \((32, 32, 48, 48, 64, 80)\)). The TCM has 5 blocks with dilation rates \((2^0, 2^1, 2^2, 2^3, 2^4)\), kernel size of 3. The model includes 4.75M parameters. It is trained to predict the magnitude mask and the phase in the spectrum domain. The model ranks the first in 2021 ICASSP DNS Challenge\footnote{\url{https://www.microsoft.com/en-us/research/uploads/prod/2020/12/Challenge_Results.pdf}}.

\subsection{Evaluation results}
\label{ssec:evaluation}


 \begin{table}[t]
\centering
\vspace{-0.2cm}
\caption{Comparison on synthetic test set.}
\begin{tabular}{ c|c|c|c}
\hline
Scheme&SIG&BAK&OVR \\
\hline
noisy &3.63&3.44& 3.1\\
TFCodec-clean&3.93 &4.46 &3.73 \\
TFNet&3.59 &4.34 &3.44 \\
proposed generative&\textbf{3.71} &\textbf{4.41} &\textbf{3.56} \\
\hline
\end{tabular}
\label{table:main_tabel_1}
\end{table}


Table \ref{table:main_tabel_1} and Table \ref{table:main_table_2} show the performance comparison on the synthetic test set and real-recorded blind test set, respectively. Compared with TFNet, it can be seen that the proposed generative framework has largely improved both the speech quality with less speech distortion by better SIG and the denoising capability by better BAK. Take TFCodec-clean which is the reconstruction from ground-truth clean codes as a reference, the BAK of the generative model is quite close to TFCodec-clean, which indicates the noise robustness by the clean codes of the speech codec. More audio samples can be found at our demo page\footnote{https://dns-gen.github.io/} to see its ability to remove the unseen noise types where TFNet has the leakage. Furthermore, the WER results from Table \ref{table:main_table_2} also show better intelligibility and reliablility of the proposed generative framework.

\begin{table}[t]
\centering
\vspace{-0.2cm}
\caption{Comparison on blind test set of DNS Challenge at ICASSP 2022. Except for WER, the higher the better.}
\begin{tabular}{ c|c|c|c|c }
\hline
Scheme&SIG&BAK&OVR&WER[\%]\\
\hline
noisy&4.14 &3.0 &3.3&39.36\\
TFNet&4.06 &4.41 &3.79&24.82\\
proposed generative&\textbf{4.27} &\textbf{4.63} &\textbf{4.09}&\textbf{17.19}\\
prefix-based generative&4.0 &4.48 &3.81&60.74\\
predict-synthesis&4.09 &4.51 &3.86&42.69\\
\hline
\end{tabular}
\label{table:main_table_2}
\end{table}

\subsection{Ablation studies}
\label{ssec:ablation}
\subsubsection{Comparison with prefix-based approach}


Here we compare our explicit-alignment scheme with the typical prefix-based conditional generative framework as shown in Fig. \ref{Fig_proposed_framework}(b). Without explicit alignment, the prefix-based method needs to learn the alignment between the condition and the clean codes. To clarify, the results of prefix-based generative framework in Table \ref{table:main_table_2} are obtained by conditioning on all noisy frames, that is non-causal in noisy condition part, but we can still see its large degradation compared with the explicit-alignment approach. The degradation is mostly caused by the errors in the predicted speech, e.g.missing/repeating/dis-ordered content which is also verified by the high WER. The proposed approach is much more robust and able to scale to longer sequences with its explicit alignment design although trained with fixed-length input.
\subsubsection{Comparison with predict-synthesis framework}
\label{sssec:NAR_approach}
To further verify the superiority of the auto-regressive (AR) generative design for low latency speech enhancement, we implement a predict-synthesis framework based on a causal non-autoregressive (NAR) transformer in which the clean codes of TF-Codec is first predicted from noisy signal and reconstruct to clean waveform with the codec decoder. From Table \ref{table:main_table_2} we can see that although the BAK is still high because of the noise robustness of the codec codes, the speech quality by SIG is much lower because of the worse temporal coherence.

\section{CONCLUSIONS}
\label{sec:conclusions}
In this paper, we propose a conditional generative framework for low-latency speech enhancement based on neural codec token generation. With explicit-alignment between noisy condition and clean codec codes and the group-VQ based single-stage generation scheme based on a decoder-only AR transformer, it achieves good noise robustness and speech temporal coherence compared with mainstream data-driven based approaches, especially on unseen noise types with complex noisy scenarios. Ablation results also show its higher accuracy on the content prediction and better scalability to long sequences than typical prefix-based generative framework and better temporal coherence against NAR-based predict-synthesis framework.

\vfill\pagebreak

\bibliographystyle{IEEEbib}
\bibliography{strings,refs}

\end{document}